# Impact of DG and survey of pricing mechanism around different states in USA

Mohsen Abedi
Washington State University, USA
Mohsen.abedi@wsu.edu

*Abstract—* *In this paper, we investigate the impact of the Distributed generation (DG) on Power Transmission Grid and pricing strategy for different geographical locations in different states, generally distributed generation (DG) on distribution feeders is known to have an impact on voltage regulation and power quality. A DG might provide voltage and power support in some cases, but most cases cause short term instability voltage, depending on the several variables, including relative DG size and location, power network and load character, so This paper also addressing to the following questions: the different between DG and conventional power station? What are penetration levels of different DG to the power grid and compliances to the IEEE relevant standards, discusses unexpected issues with a special emphasis on power protection coordination problems, potential advantage and disadvantage to using DG while Distributed Generation (DG) allows collection of energy from many sources and may provide lower environmental pollution then we will discuss what would be the main Factors affecting Electricity Prices of DG units at different states.*

## I. Introduction

As more demanding of electric power are in place day by day in the power market. it is going to be difficult for major utility to provide economical following by the technical solution to their network and customers, so new reliable smaller and expandable power sources (DG) defined variable energy source up to 10 MW are in demand by considering of some conditions and restriction plus computational requirements, those can set into or out of the power network with minimum required time of synchronization and adjustment which is called fast setup, but even these type of power sources (DG) would be causing some unexpected issues that might need to be considered and limited or resolve in order to maximize the power performance and efficiency. Impacting of Distributed generation on electric power network are mostly faced in the low voltage networks, such as:

- voltage adjustment and fluctuation, voltage profile if it is using different optimization techniques.
- Coordination of protection devices in distribution network.
- Transient disturbances, power flow direction and magnitude
- Protection and safety problems, overcurrent Protection scheme and Overcurrent Device Coordination.
- Excessive harmonic content which can cause severe issue.
- The Impact of Distributed Generation systems in the load forecasting and price forecasting.
- problems on the customers and power quality.
- Integration in electric power distribution systems on fault location methods.
- Influence of Distributed Generation on Congestion and LMP in Competitive Electricity Market.
- Power network stability issue.
- required own pricing strategy in different states.

Increased penetration of distributed generators (DG) based on wind and PV solar systems have brought new challenges to power systems, such as, ensuring system stability, voltage regulation and power quality within standard limits. [1-2]. Also, if generated power by DG system becomes more than demanded load power then excess power flows in the reverse direction towards the main grid past to the substation transformer. The reverse power flow will be causing the potential result such as voltage rising on main feeder [1].

It is depending on the distribution network operating characteristics and the DG characteristics. DG can be beneficial if it meets at least the basic requirements of the system operating philosophy and feeder design. The effect of DG on power quality depends on type of DG, its interfaces with the utility system, the size of DG unit, the total capacity of the DG relative to the system, size of generation relative to a load at the interconnection point and feeder voltage regulation practice [3]. DG might be causing the harmonic distortion (THDI)in the power network, it is depending to the type of DG and power converter systems, generally issues arise when different type of distributed generators and technologies are tied to the distribution network, as there are many DG technologies utilize renewable energy due to growing attention to air pollution and greenhouse effects. More than 12 million DG units are installed across the United

States today, with a total capacity over 200 GW. In 2003, these units generated approximately 250,000 GWh, over 99% of these units are small emergency reciprocating engine generators or photovoltaic systems, installed with inverters that do not feed electricity directly into the distribution grid, this large number of smaller machines represents a relatively small fraction of the total installed capacity, under locational marginal pricing, the price of energy at any location and states in a network is equal to the marginal cost of supplying an increment of load at that location, EPACT Section 1252 contains standards for smart metering and time-based pricing which are generally considered to be important "enabling mechanisms" for consideration of investments in DG by consumers and electric power companies for each states. [4], so pricing mechanism are varying in states depending on rules and regulations made by states, fuel cost counted by available natural resources and load tariff or distance to the network, in this paper we are discussing on all relevant parameters whose might be considered on pricing mechanism between states.

## II. POWER QUALITY & PROTECTION ISSUES BY DG

*A. Voltage issue*

The control and operational strategies of the DG (both static and dynamic) are different from conventional power systems which The voltage gradient along the distribution feeders are of concern, so The voltage issue comes as common and most important matter that need to be address in order to have reliable power source if it is provided by DG in power network or DG power combined to the power grid, some DG voltages control are static voltage regulators (SVR) (1547A) while some others are load tap changers (LTC), the LTC voltage control units are not reliable voltage regulation because they comes with lack of the ability to produce reactive power or controlling of the power factor, so might result the voltage collapse which it occurs when an electric system does not have adequate reactive support to maintain voltage stability (figure 2.1). Voltage collapse may result in outage of system elements and may include interruption in service to customers.

In some cases, the voltage regulation is also enhanced by switched capacitor banks placed along the feeder, Line Drop Compensators (LDC) are seldom used except in rural areas where load density is low, or in substations where each feeder is fitted with an independent voltage regulator, also Load models can generally be classified into two types: static and dynamic models.

Static load models in the load flow studies are considered to be of major importance because only active and reactive steady-state powers are relevant at bus voltage level. One form of voltage control is through reactive power compensation. Shunt capacitor banks for generating reactive power, installed in the substation or along the feeder can boost the voltage. By consuming the reactive power. Loads and long overhead transmission lines do consume reactive power but this depends on the type of loads connected at the end of the line. Voltage control needs reactive power either inductive from the loads or capacitive from the lines [8], the DGs causes a rise in voltage; similarly, a drop in the generated power causes a drop-in voltage. These actions bring fluctuations referred to as voltage swells and sags usually above stipulated allowed values of ±5% of the nominal values (EN50160).

The PV curve is useful to calculate the voltage collapse due to required reactive power needed to prevent the voltage collapse.

$$P = P_0 (a_0 + a_1 V + a_2 V^2)$$
$$Q = Q_0 (b_0 + b_1 V + b_2 V^2)$$

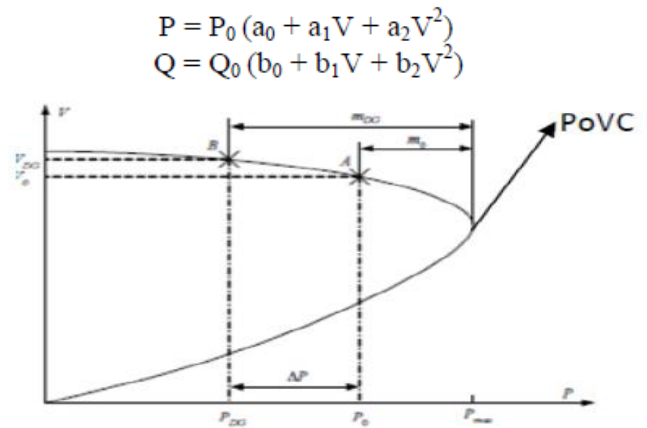

Figure 2.1 PV curve

As example, Lack of the coordination between DG with the power grid while there is line shortage to the ground and causing to trip the main feeder breaker that would cause to increase voltage on those remain customers feeding by DG and isolated from main feeders, depended to nature of fault, it cause to increase the voltage on one phase or two phases or drop the voltage in all three phases lines feeding from DG to the local customers, duration of the over voltage could be in milliseconds which still is harmful because Temporary overvoltage oscillatory overvoltage persisting for many cycles even to seconds, depended to the power line distance [9].

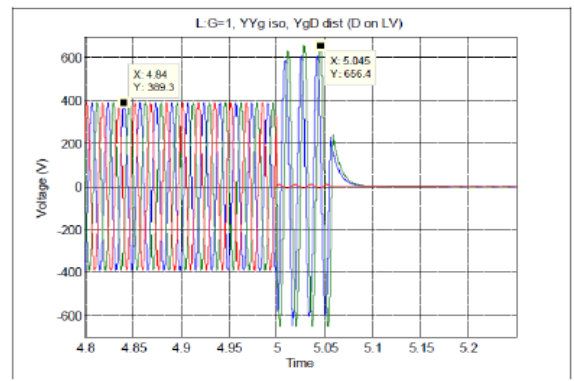

Typical response seen when load is suddenly rejected and load:gen ratio changes
Figure 2.2 voltage flicking

Also, there is chance of Voltage flicker which it happens while input energy is vary or results from intermittent generation from wind turbines and photovoltaic sources, or connection and disconnection of induction generators from the network.

Most cases DG could effect on Line drop compensation (LDC) in distribution line Which result lower voltage at end of the line than desired, which it is required interface and proper communication with all set points in distribution line and Power grid while some DG small unit doesn't have this facility for communication and interfacing, so IEEE 1547 recommended smart grid to using of End-of-line monitoring (EOL) system combined to the network monitoring system such SCADA, in order to minimizing the issue.

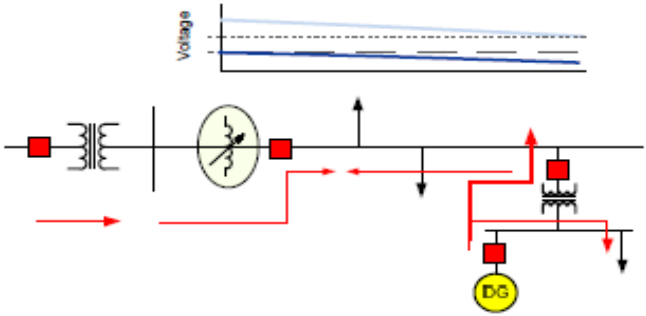

Figure 2.3: LDC regulation distracted by DG

*B. Harmonic issue*

Harmonic distortion is causing by the non-linearity of equipment's such as power converters, transformer, rotating machines, there are forms of DG such as photovoltaic (PV) double fed induction generator (DFIG) wind turbine, and fuel cells are tied to the power grid and function by power electronic inverter based on IGBT or SCR bridge which these type generally causing back feed unwanted current harmonics THDV (2) or even THDI (1) to the power grid, these values are vary depended to size of the load and load power factor and other relevant parameters[10].

$$THD_V = \frac{\sqrt{V_2^2 + V_3^2 + V_4^2 + \cdots + V_n^2}}{V_1} * 100\% = \frac{\sqrt{\sum_{k=2}^{n} V_k^2}}{V_1} * 100\% \quad (2)$$

$$THD_I = \frac{\sqrt{I_2^2 + I_3^2 + I_4^2 + \cdots + I_n^2}}{I_1} * 100\% = \frac{\sqrt{\sum_{k=2}^{n} I_k^2}}{I_1} * 100\% \quad (3)$$

$$V_{rms} = V_{1,rms}\sqrt{1 + \left(\frac{THD_V}{100}\right)^2} \quad (4) \qquad I_{rms} = I_{1,rms}\sqrt{1 + \left(\frac{THD_I}{100}\right)^2} \quad (5)$$

The DG with the PV module can be characterized to using an electrical equivalent circuit of a solar cell that contains a current source anti-parallel with a diode, a shunt resistance and a series resistance as shown in figure 2.4.

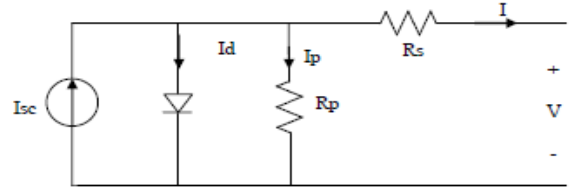

Figure 2.4: equivalent circuit of solar cell

$$I_{sc} = I_{scR} \frac{G}{G_R}[1 + \alpha_T(T_c - T_{cR})] \quad (6)$$

ISCR (6) is short circuit static current and it is representing of current harmonic generated by diode bridge.
As it is shown on figure 2.4 the PV panel considered a DC source so it is required to convert DC to AC by inverter bridge by using PWM topology then modulated to stimulated sinewave 60 HZ by RLC filter which it won't be purely sinewave so it would be easily disturbed by nonlinear load then generate back feed harmonic, this unwanted harmonic would be influencing on power control and metering systems in common grid then disturbing the nodes LMP to be accurate and pricing of congestion in vary locations due to miss-calculation of actual power flow .

So, it is required a method for Harmonic measurement of Real Power grid signals with frequency drift using instruments along with internally generated reference frequency, so there is economic impact of the power quality by having unwanted harmonics generated induced by DG unit which the fundamental RMS values of voltages and currents reading (4), (5) and power factor by metering units would be change and not accurate, so metering relevant to control devices units picking up wrong values. IEEE 519 specified the quid line for Harmonic control and compensation of static power conversion in order to minimized the harmonic distortion but still DC/AC or AC/DC.

Convertors based on SCR or IGBT (figure 2.5) bridge are main source of harmonic that need to be filter before injecting the power to the grid.

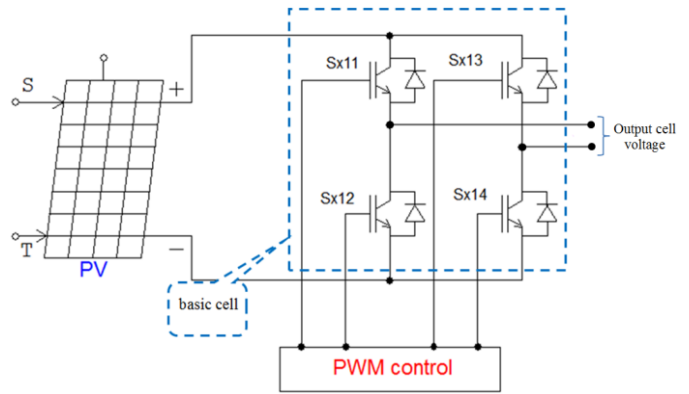

Figure 2.5: PV panel and inverter bridge.

There is Limitation of dc injection following by (IEEE® Standard 1547-2003 4.3.1) which The DG and its interconnection system shall not inject dc current greater than 0.5 percent of the full rated output current at the point of DG connection.

*C. Current protection issue*

Overcurrent protection schemes for radial distribution systems are designed based on the available short circuit ratios, maximum load currents, system voltage and insulation levels. The addition of generation on the feeder results in altered currents flowing in various parts of the feeder for faults at different points on the feeder. The primary concerns for DG interconnection are typically sympathetic tripping issues, failure of fuse-saving schemes and reduction of reach - potentially resulting in undetected faults, Sympathetic tripping occurs when a protective device operates unnecessarily for faults in other protection zones. This can occur with distributed generation due to unexpected fault contributions from the DG, an example of how sympathetic tripping might occur is shown in Fig. 2.6. The relays at breaker "A" and the recloser are not directional. Thus, sufficient fault current infeed from the distributed generation would cause either of these devices to operate in "sympathy" with" B" which actually sees the fault. [5]

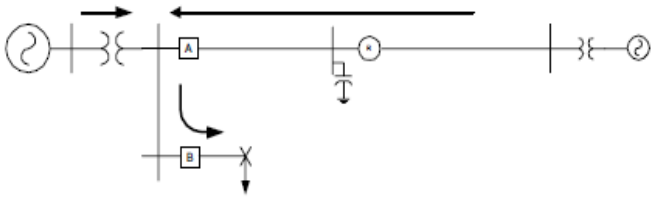

Figure 2.6: Schematic illustrating sympathetic tripping

Another potential overcurrent protection disruption is de-sensitization of the feeder overcurrent protection, also referred to as the reduction of reach of the feeder protective devices. "Reach" refers to the distance downline of the protective device to which the device can detect a fault. Without DG, only the utility source feeds a fault, and the currents flowing into the fault on a radial circuit are easily calculated. Utility protection engineers typically coordinate protective devices by setting the pickup current so that the device will operate for a selected smallest minimum fault current expected, which
correlates correspond to the highest impedance fault to be detected. On radial feeders, up to 85% of faults are initially transient and may be self-cleared within a few cycles.

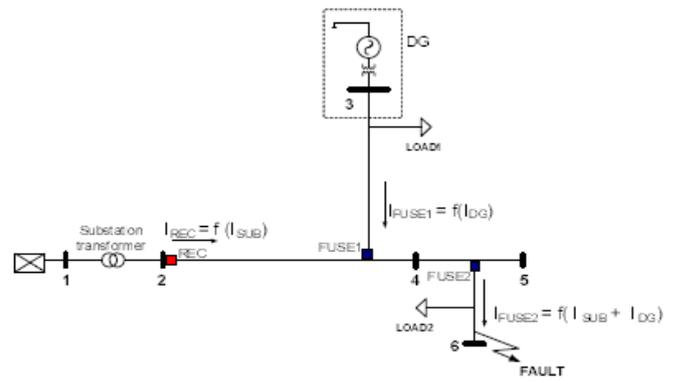

Figure 2.7: Typical radial feeder with fuse and recloser

According to the figure 2.6, as example, there are Three potential issues concerning downstream fuses such as fuse fatigue, nuisance fuse blowing, and fuse miss-operation. The main cause of the issues is due to the redistribution of the fault current on the feeder between the DG and the substation. Fuse fatigue arises when the fuse begins to melt before the recloser's fast operation, Nuisance fuse blowing occurs when the fuse blows prior to the recloser's fast operation. The result of nuisance fuse blowing is a permanent outage on the faulted lateral [6], If there is a fault as shown in Fig. 2.6, FUSE1 may operate incorrectly and isolate the un-faulted lateral. This issue is known as fuse miss-operation, Fuse miss-operation is caused by the infeed current from the DG to a fault that is located outside the lateral protected by the fuse. These issues may result in unwarranted permanent outage and islanding during a temporary fault. Studies indicated that fuse fatigue and nuisance fuse blowing were dependent on the total penetration level of the DGs and the locations while fuse miss-operation was dependent on the number of DGs at different locations on the feeder. Also, it is recommended to avoiding any island operation result by DG because it could be causing miss operation of main fuse of feeder to properly protect that specific area resulted by Island, also Over current relay setting on a feeder must be redefined accordance to the added DG to the line. If the utility has a policy that the DG does not supply current to the utility (reverse power flow), a directional power flow relay may be used to detect this condition for a predetermined amount of time. So Then DG will only be permitted to deliver power to the local loads in a "peak-shaving" type application

### III. IMPACT OF DG ON POWER TRANSMISSION NETWORK

In recent years, distribution generation (DG) has been one of the most attractive research areas in the field of power generations. In general, DG can be defined as electric power generation within distribution networks or to the customer side of the network. Hence, so it is essential to study the impacts of DG in the transmission network, problems are generally, classified as static and dynamic. Static transmission network expansion planning (STNEP) method determines new transmission facilities needed to meet the system requirement for a specific planning horizon while Dynamic TNEP (Transmission network expansion planning problem) is timing based method. Also, Transmission investment cost is more when generating station is far away from load centers. therefore, impact of large scale integration of DG in TNEP problem is presented and need to be considered [7].

Generally, issues involved in the exploration of distributed generation and influences on Transmission grid such: 1) how to simply define a characteristic number to quantify the amount of distribution from some combination of the fraction of power from distributed generation and the fraction of nodes with distributed generation Figure 2.8.

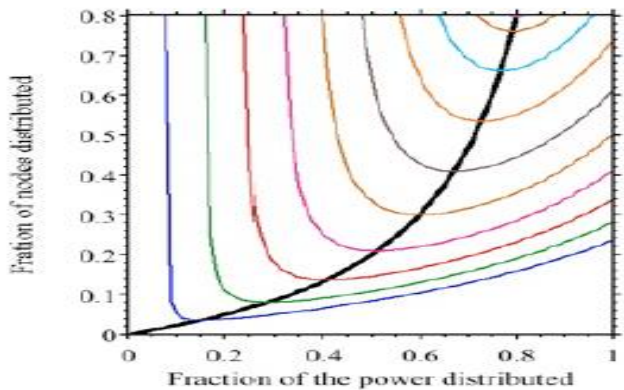

Figure 2.8: Distribution degree as function of distribution power fraction and distributed node fraction.

2) the impact of the reliability of the distributed generation, 3) economic upgrade models for the various types of generation. 4) dispatch models for the distributed generation and combining with the existing dispatch models. 5) DG would be impacting on the coordination of protection transmission devices through relay disorganization or reduction in reach of distance relays, incessant and incorrect tripping and islanding of power network, for example the distance relays are set to operate in specific time for any faults occurring within a predefined Zone of transmission line, Due to presence of DG, a distance relays may not operate according to the define zone settings, also this might be results on any other dispatch sensors those are using to configuring the network either for economy dispatch or protection, again DG units if they are not configured accordance to the network function then it might disturb all zone setting [6], table 1.

| Zones | Relay settings (% of line length) | Distance relay operating range | | | |
|---|---|---|---|---|---|
| | | 3 phase fault | | 1 phase fault | |
| | | w/o DG | with DG | w/o DG | with DG |
| Zone 1 | 40 | 40 | 40 | 39 | 39 |
| Zone 2 | 80 | 79 | 67 | 79 | 74 |
| Zone 3 | 115 | 100 | 91 | 100 | 100 |

Table 1: operating zones of distance relay or sensors with DG and without DG

So, it is recommended to follow some standards and regulations such as IEEE 1547 and 2030 Standards can be used for some of those interacted issues regarding of Distributed Generators Interconnection and Interoperability with the power Grid and transmission lines. For example, the following table:2 from IEEE 1547 using as DG frequency reference or rejection point from main Power grid that is required to integrating of DG units with the utility Power.

| DR size | Frequency range (Hz) | Clearing time(s)[a] |
|---|---|---|
| ≤ 30 kW | > 60.5 | 0.16 |
| | < 59.3 | 0.16 |
| > 30 kW | > 60.5 | 0.16 |
| | < {59.8 – 57.0} (adjustable set point) | Adjustable 0.16 to 300 |
| | < 57.0 | 0.16 |

Table 2: IEEE 1547 Trip Values for frequency reference between DG and Power Grid.

## IV. INFLUENCE OF DG ON CONGESTION AND LMP IN COMPETITIVE ELECTRICITY MARKET

Distributed generators are incremental power sources in the power grid and they are widely growing in the power network that for sure we need to consider their influence on marketing and pricing mechanism, also the unexpected congestion arise by using DG units in power transmission line. So, it is depended to DG location then the DG power source can relief the transmission congestion and corresponding locational marginal Price (LMP) which by knowing the right topology to use the DG at the right location with proper security and communication system then we will be able to have some benefit of using the DG units otherwise it will be causing distortion in our pricing and marketing strategy. If the DG installed close to the load that is fast act reliable source without causing line congestion but there is no such this promise that we can manage DG location to be selected at most desire location, so still it is required management of power flow plus implementing of operation strategy and optimization by knowing the DG size and location in network.

Recently, the congestion management schemes involving the deployment of FACTS devices and DG have been envisaged as promising options, traditional concept of load curtailment and Price area congestion management is considered and a modified approach to congestion management based on locating DG in an OPF based market has been proposed. In other hand in problem formulation. the objective function is formulated as quadric benefit curve submitted by buyers (DISCO) minus quadratic bid curve supplied by sellers (GENCO) minus quadratic cost function by the DG owner [11]

$$\max \sum_{i=1}^{N} \left( B_i(P_{Di}) - C_i(P_{Gi}) \right) - C(P_{DGi}) \quad (7)$$

$$P_i = P_{Gi} + P_{DGi} - P_{Di} = v_i \sum_{j=1}^{N} \left[ v_j \left\{ G_{ij} \cos(\delta_i - \delta_j) + B_{ij} \sin(\delta_i - \delta_j) \right\} \right] \quad (8)$$

If we just considered the active power of DG injected to the grid the Lagrange for social welfare model can be as below:

$$L = \sum_{i=1}^{N} \left( C_i(P_{Gi}) - B_i(P_{Di}) + C(P_{DGi}) \right) + \sum_{i=1}^{N} \lambda_{Pi} \left( P_i - P_{Gi} - P_{DGi} + P_{Di} \right) + \sum_{i=1}^{N} \mu_{Pi}^{\min} \left( P_{Gi}^{\min} - P_{Gi} \right) + \sum_{i=1}^{N} \mu_{Pi}^{\max} \left( P_{Gi} - P_{Gi}^{\max} \right) \quad (9)$$

If the reactive power influence by DG ignored then LMP would be reduced at the node where DG placed, also this placement would be effect on other nodes in beside which in general is benefit for all customers tied to these nodes,

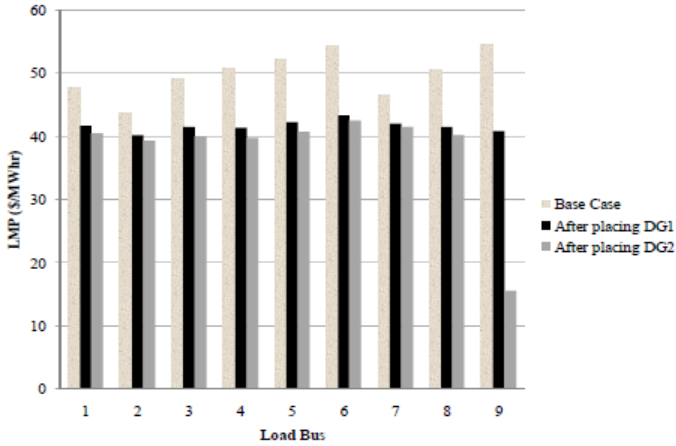

Figure 3.1: Load Bus LMP corresponding in three different cases

According to the figure 3.1 resulting of 9 buses network, the LMP reduced from base case to the case with DG, so by having the DG1 and DG2 placement then line congestion also improved, the cheaper unit would be higher penetration as net social welfare, still size of DG is most important factor, the optimal dispatch from DG have been founded to reduce the congestion rent and shadow prices associated with the line flow, so DG with the minimum incremental cost is founded better performance in terms of alleviating congestion in power network [12] , [11].

## V. REGULATORY FRAMEWORK FOR DG AT DIFFERENT STATES OF USA

One major DG regulatory issue is the recovery of stranded costs by utilities.35 In 2002, these costs were "being covered via a fee on future electricity sales or imposed as a charge on those individuals or businesses exiting the utility system."36 In 2002, Massachusetts was the only state to address this issue. Massachusetts recommended "that if a proposed DG system exceeds 50 percent efficiency then the withdrawal fee is waived."37 A second major regulatory issue is interconnection with the grid. Interconnection requirements vary from utility to utility and from state to state. "Energy providers seeking to retail electricity, whether DG or the grid, must be licensed to do so. The license process varies by state [13].

Depending on state regulations, some customers may be eligible for net metering. If eligible, net metering effectively allows a customer to sell excess generation back to the grid at the same retail local price as the customer pays for power from the grid during other periods. With this financial incentive, the economics for small commercial and residential DG installations, especially those based on renewable energy technologies, are substantially enhanced.

| State | License requirements Apply to | Eligible fuel |
|---|---|---|
| AZ | Any company supplying, marketing or any competitive electric services pursuant | Renewable and Cogeneration |
| CA | Any company that sells electricity to residential and small commercial customers (maximum peak demand less than 20 kW) | Solar & wind |
| CT | Any entity including an electric aggregator or participating municipal electric utility that provides electric generation services to end- use customers | Solar, wind, hydro, fuel cell, sustainable biomass |
| MA | A retail seller of electricity whose price to a consumer is not regulated and which is not a municipal light department or a distribution company. | Renewable & cogeneration |
| MD | Suppliers/brokers/billers offering retail electric supply of electric generation services in Maryland | Solar only |
| VT | Any person or company under the jurisdiction of the Public Service Board that is engaged in the sale of electricity to retail consumers in Vermont. | Solar, wind, fuel cells using renewable fuel, anaerobic digestion |

Table 2: example of some states policy for DG license

As table 2 indicated to the license requirement applying for some different states policy and restriction as example that considering states resource available for different type DG units and market policy.

The Renewables Portfolio Standard ("RPS") is one of the most ambitious renewable energy standards in USA, The RPS program requires investor-owned utilities, electric service providers, and community choice aggregators to increase procurement from eligible renewable energy resources to 33 percent of total procurement by 2020. The California Public Utilities Commission and the California Energy Commission jointly implement the RPS program. The CPUC's responsibilities include (1) determining annual procurement targets and enforcing compliance; (2) reviewing and approving each IOU's renewable energy procurement plan; (3) reviewing IOU contracts for RPS-eligible energy; (4) establishing the standard terms and conditions used by IOUs in their contracts for eligible renewable energy; (5) calculating market price referents (MPRs) for non-renewable energy that serve as benchmarks for the price of renewable energy, The CEC's responsibilities include (a) adopting regulations specifying procedures for enforcement of the RPS for publicly owned utilities; (b) certifying and verifying eligible renewable energy resources procured by publicly owned utilities and to monitor their compliance with the RPS; (c) continuing to certify and verify RPS procurements by retail sellers, and (d) referring failures of a publicly owned utility to comply to the Air Resources Board. [13]

As it is indicated the renewable portfolio standard (RPS) is one of the most common state-level renewable energy policies in the USA today. Gradually every single state generated their own policy to get DG system in their market so then it was requiring the regulated mechanism of TRIF as pricing base on those implemented standards, also in order to encourage utilities to invest in energy efficiency programs like DG, some states mandate energy and emissions reductions.

For example, Massachusetts is a leading state with a long, successful record of implementing energy efficiency programs for all customer sectors which the state created an aggressive funding mechanism and required electric utilities to provide energy efficiency programs during its restructuring of the industry in 1997, so DG was one the most Incentives program in Massachusetts that brought the attention to use the natural gas utilities for DG units.

Gradually Massachusetts' implemented the Solar Carve-Out a market-based incentive program to support residential, commercial, public, and non-profit entities in developing 400 MW of solar photovoltaic across Massachusetts.

Massachusetts' Solar Carve-Out provides a means for SRECs to be created and verified, and allows electric suppliers to buy these certificates in order to meet their solar RPS requirements. New Jersey also regulates DG through both consumer and utility programs. Indeed, New Jersey's renewable portfolio standard is one of the most aggressive in the United States, it is requiring each supplier/provider serving retail customers in the state to procure 22.5 percent of the electricity it sells in New Jersey from qualifying renewables by 2021 In addition, the standard also contains a separate solar specific provision which requires suppliers and providers to procure at least 4.1 percent of sales from qualifying solar electric generation facilities by 2028.

In some other states such as Ohio's Energy Efficiency Portfolio Standards, electric utilities are required to implement energy efficiency and peak demand reduction programs that result in a cumulative electricity savings of 22 percent by the end of 2025, with specific annual benchmarks, in addition, utilities must reduce peak demand by 0.75 percent annually through 2018, then the legislature must make recommendations for future peak demand reduction targets, In order to meet the targets, utilities may implement demand-response or customer-sited programs, or transmission and distribution infrastructure improvements.[13]

Gradually the most states motivated to Support the DG renewable energy units according to their available natural resource for DG units, so they started establishing the renewable portfolio standards, by 2015, twenty-nine states and the District of Columbia had renewable portfolio standards.

## VI. PRICING MECHAMNISM AROUND DIFERENT STATES IN USA

The easiest common tool to track electrical output from distributed Generators (DG) units such as solar generators to compensate distributed generation owners for this output is a billing arrangement known as "net metering." Since 2001, net metering has been available to utility customers in a majority of states, the passage of the 2005 Energy Policy Act, however, catalysed distributed generation under net metering by offering favourable tax treatment to individuals installing solar generators and by encouraging state adoption of net metering policies, as a result, net metering has grown significantly in the wake of the 2005 Act. By 2014, net metered residential solar customers of some other type DG units had collective generation capacity of 2,850.780 MW123 and accounted for 19.161% of just total photovoltaic capacity in the United States.

However, despite near ubiquitous adoption of net metering by states, the policies themselves differ, often substantially, between jurisdictions, EPACT Section 1252 contains standards for smart metering and time-based pricing.

state net metering programs differ in how they compensate customer-sited generation, Thirty-four net metering jurisdictions credit customers for generation at the retail rate.

many states offer a combination of rates, this combination typically credits monthly excess generation that is "carried-over" to future billing cycles at a retail rate, but credits annual net excess, when utilities and net metered customers "zero-out" generation and consumption from the past 12 months, at an avoided cost rate.

In Wisconsin, for example, utility avoided costs rates are $0.030 to $0.040 per-kWh, while retail rates range between $0.110 and $0.140 per-kWh.

Retail rates in Kansas, which are some of the nation's highest, can reach $0.190 per-kWh, while utility buy-back rates for excess generation can be as low as $0.013 per-kWh, or just 7% of the retail price.

Fourteen states and the District of Columbia credit customer excess generation at the retail rate without expiration.

In Ohio, the Public Utilities Commission of Ohio recently decided that customers with distributed generation systems are entitled to the "full value" of electricity they sent to the grid, which they define as generation and capacity charges.

A second variation among net metering policies is how long a customer's monthly excess generation may be "carried-over" to future billing cycles and used to offset electricity consumption.

Thirteen jurisdictions offer customers some variation of indefinite carry-over, though most state policies limit how long excess generation may be applied in subsequent billing periods.

Twenty-two states limit the available carryover to 12 months, Enabling the carryover of excess generation, even if limited to twelve months, leads to very low electricity bills for customers that own large photovoltaic systems.

Third, nearly all jurisdictions place a cap on the maximum size of any individual net metered generator, these limits can range from relatively restrictive, like the 10-kW ceiling in Georgia, to more generous limits like the 80 MW cap in New Mexico.

Fourth, twenty-four jurisdictions set aggregate capacity limits that constrain the total amount of net metered generation permissibly installed within a state or utility service area, typically expressed as a percentage of the yearly or historical peak demand for electricity, aggregate limits commonly fall between 0.2% and 9.0%, For example, Vermont has an aggregate capacity limit at 15% of the state's peak demand, many states, however, do not have an aggregate capacity limit. [14]

As example, Hawaii, adopted new tariffs and lowered compensation for new customers, which had been at the $0.298 per-kWh retail rate, to between $0.150-0.280 per-kWh, the "avoided cost" rate, Hawaii's new policy, which replaces net metering, offers distributed generation owners a choice between a "grid-supply tariff," which reduces compensation to avoided cost rates.

In January 2015, Nevada's state utility commission upheld a change to state net metering that decreases the credit offered for energy sold back to the grid from the retail rate of $0.11 per-kWh to $0.026 per-kWh over the next four years.

A. *The Utility "Purchase Obligation"*

In most cases, electric utilities are required by federal law to provide service to customers who choose to install DG. Pursuant to rules authorized by the Public Utility Regulatory Policy Act of 1978 (PURPA) and promulgated by the Federal Energy Regulatory Commission (FERC), utilities must offer to sell electric energy to and purchase electric energy from "qualifying small power production facilities" and "qualifying cogeneration facilities" at rates that are just and reasonable to the utility's customers and in the public interest, and non-discriminatory toward qualifying facilities.

Under federal law, a "small power production facility" is a generating facility of 80 MW or less whose primary energy source is renewable (hydro, wind, or solar), biomass, waste, or geothermal resources in order to be considered a qualifying facility (QF), a small power production facility must meet all of the requirements in FERC rules for size, fuel use, and certification, facilities with a rated capacity of 20 MW or less do not have non-discriminatory access to markets, and a rebuttable presumption that facilities greater than 20 MW capacity do have non-discriminatory access in five of the seven U.S. wholesale electricity markets: the Midcontinent ISO, PJM Interconnection, New York ISO, ISO New England, and Electric Reliability Council of Texas.

Furthermore, the FERC rules require each utility to offer standard rates for purchases from all QFs with a design capacity of 100 kilowatts (kW) or less, FERC gives utilities discretion on whether to offer standard rates or to individually negotiate rates for purchases from QFs larger than 100-kW capacity, but state laws and regulations may further limit that discretion.

B. *Eligible Technologies.*

Despite the fact that the PURPA ratemaking standard required consideration of NEM for all forms of DG, in practice most states and utilities have more narrowly specified the technologies that are eligible for NEM tariffs. Fossil fuelled generators, including CHP systems and diesel generators, are eligible for NEM only in a small number of states, almost all NEM tariffs include a maximum limit on the size of eligible DG systems.

C. *Allocation of Renewable Energy Credits (RECs).*

State policies and individual utility tariffs also vary in the ways they treat REC ownership under NEM arrangements. Most state policies grant ownership of any RECs created under a NEM tariff to the customer, or do not specify who owns the RECs . A few states (e.g., New Mexico) grant REC ownership to the utility or require sharing of the RECs between the customer and the utility. Where REC ownership is not specified in state policy, it may or may not be specified in an individual utility's tariff. Some states also require customers to transfer RECs to the utility if state or utility subsidies were used to support the installation of the system.

D. *Meter Aggregation.*

Nearly 20 states have adopted policies that allow for the aggregation of multiple meters under a NEM tariff. States vary in what they allow, figure 5.1 showing the Net metering uptake by 2011.

Generally speaking, the output of a single generator is allocated to all of the participating meters
and netted against the consumption measured on those meters as with other NEM tariffs'

This sort of arrangement is sometimes referred to as "group," "community,", "neighbourhood," and "virtual" NEM or aggregation.

When a utility offers a feed-In-Tariff (FIT), it essentially offers to enter into a long-term power purchase agreement, under standard (non-negotiable) terms and conditions, with any customer who meets specified eligibility criteria. a FIT offers the customer a price that exceeds the utility's avoided costs of purchasing unspecified energy and capacity, Most FITs are structured in such a way that the utility agrees to pay the customer a fixed price for every kWh the customer generates over the duration of the contract.

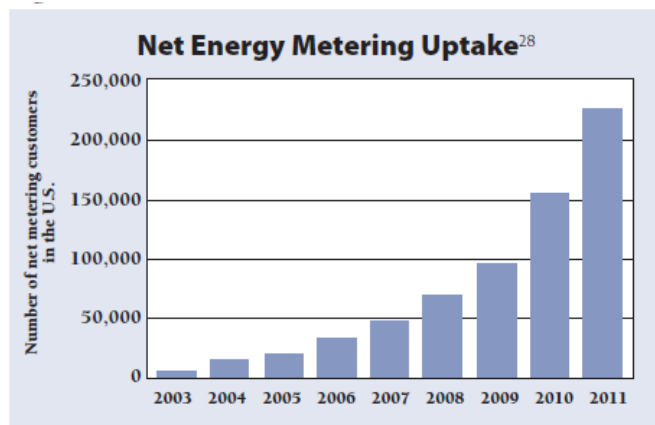

Figure 5.1: Net Metering Un take by 2011

As with NEM policies, FIT policies can be established either by a state legislature or by a state PUC and may apply to different combinations of IOUs, electric cooperatives, and municipal electric utilities, the prices paid to customers under a FIT can be determined through either of two procedural methods:

**1**- The most common method historically has been for the utility or the PUC to set FIT prices through an administrative process, such as a normal tariff proceeding. In some of these jurisdictions, FIT prices are based primarily on estimating the generator's costs.

**2**-An entirely different procedural method for setting FIT prices is to use a competitive procurement process, with this kind of method, the utility establishes all of the terms and conditions of the FIT except the price and then solicits price bids from potential participants through a Request for Proposals or a reverse auction. [15]

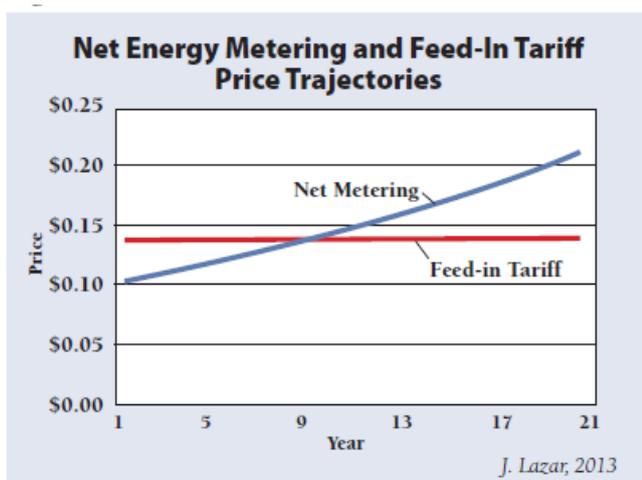

Figure 5.2: Net Energy and FIT pricing path.

However, the more immediate task faced by regulators today is to adapt the NEM and FIT mechanisms to ensure fair compensation on all sides, also the pricing mechanism are varying in states depending on rules and regulations made by states, fuel cost counted by available natural resources and load tariff or distance to the network those we already discussed on this paper about all relevant parameters whose might be considered on pricing mechanism in states.